\documentclass[preprintnumbers,showpacs,aps,amsmath,amssymb,superscriptaddress,nofootinbib,prl,twocolumn,10pt]{revtex4-1}
\usepackage{epsfig}
 \usepackage{euscript}
\usepackage{amssymb}
\usepackage{amsfonts}
\usepackage{amsbsy}
\usepackage{amsmath}
\usepackage{epsfig}
\usepackage{amsthm}
\usepackage{amscd}
\usepackage{amstext}
\usepackage{verbatim}
\usepackage{amsfonts}
\usepackage{graphicx}
\usepackage{epsfig}
\usepackage{eepic}
\usepackage{amsmath}
\usepackage{amssymb}
\usepackage{enumerate}
\usepackage{color}
\usepackage{dcolumn}
\usepackage{bm}
\usepackage{ulem}

\def\bc{\begin{center}}

\def\ec{\end{center}}
\def\be{\begin{eqnarray}}
\def\ee{\end{eqnarray}}
\definecolor{dyellow}{rgb}{1.,0.8,.0}
\definecolor{myblue}{rgb}{.1,.1,.7}
\definecolor{dcyan}{rgb}{.0,.6,.6}
\definecolor{dmagenta}{rgb}{0.6,0.0,0.6}
\definecolor{brown}{rgb}{0.6,0.2,0.}
\definecolor{darkblue}{rgb}{.0,.0,0.5}
\definecolor{darkred}{rgb}{0.75,0.0,0.0}
\definecolor{orange}{rgb}{1.,.6,.0}
\definecolor{dorange}{rgb}{0.8,.4,.0}
\definecolor{darkgreen}{rgb}{0.0,0.6,0.0}
\definecolor{purple}{rgb}{.4,.0,.4}
\definecolor{lightgrey}{rgb}{0.7, 0.7, 0.7}
\definecolor{grey}{rgb}{0.4, 0.4, 0.4}

\def\pa{\partial}

\newcommand{\nc}{\newcommand}
\nc{\rnc}{\renewcommand} \nc{\ket}[1]{\left | \, #1 \right \rangle}
\nc{\bra}[1]{\left \langle #1 \, \right |}
\nc{\ua}{\uparrow} \nc{\da}{\downarrow}

\nc{\braket}[2]{\langle\, #1\,|\,#2\,\rangle}
\nc{\half}{\frac{1}{2}}

\nc{\prj}{\mathcal{P}} \nc{\hilb}{\mathcal{H}}
\nc{\pth}{\mathcal{C}} \nc{\inprod}[2]{\braket{#1}{#2}}
\nc{\upket}{\ket{\uparrow}} \nc{\downket}{\ket{\downarrow}}
\nc{\upbra}{\bra{\uparrow}} \nc{\downbra}{\bra{\downarrow}}

\begin{document}
\title{A thermal quench induces spatial inhomogeneities in a holographic superconductor}

\author{Antonio M. Garc\'{\i}a-Garc\'{\i}a}
\affiliation{Cavendish Laboratory, University of Cambridge, JJ Thomson Avenue, Cambridge, CB3 0HE, UK}
\affiliation{CFIF, Instituto Superior T\'ecnico,
Universidade T\'ecnica de Lisboa, Av. Rovisco Pais, 1049-001 Lisboa, Portugal}
\author{Hua Bi Zeng}
\affiliation{CFIF, Instituto Superior T\'ecnico,
Universidade T\'ecnica de Lisboa, Av. Rovisco Pais, 1049-001 Lisboa, Portugal}
\affiliation{School of Mathematics and Physics, Bohai University
JinZhou 121000, China}
\author{Hai-Qing Zhang}
\affiliation{CFIF, Instituto Superior T\'ecnico,
Universidade T\'ecnica de Lisboa, Av. Rovisco Pais, 1049-001 Lisboa, Portugal}
\begin{abstract}
Holographic duality is a powerful tool to investigate the far-from equilibrium dynamics of superfluids and other phases of quantum matter. For technical reasons it is usually assumed that, after a quench, the far-from equilibrium fields are still spatially uniform.
Here we relax this assumption and study the time evolution of a holographic superconductor after a temperature quench but allowing spatial variations of the order parameter. Even though the initial state and the quench are spatially uniform we show the order parameter develops spatial oscillations with an amplitude that increases with time until it reaches a stationary value. The free energy of these inhomogeneous solutions is lower than that of the homogeneous ones. Therefore the former corresponds to the physical configuration that could be observed experimentally.
\end{abstract}
\maketitle
\pagebreak
Most physical processes occur under non-equilibrium conditions. Small deviations from equilibrium are
well understood in the framework of linear response theory. However the description of the dynamics beyond linear response is still one of the most challenging problems in theoretical physics. Recent experimental advances in the study of the far-from equilibrium dynamics after a quench are opening new research avenues in condensed matter \cite{11,12,10} and cold atom physics \cite{becexp}.
A typical example is the study of the spontaneous generation of defects \cite{becexp} in a Bose gas after a temperature quench across the superfluid transition which is qualitatively described by the Kibble-Zurek mechanism \cite{zurek}.

More quantitative theoretical results are known \cite{emil} in the more tractable problem of the dynamics of a zero dimensional mean-field superconductor after a quantum quench. An analytical study  \cite{emil} of the far-from equilibrium time evolution of a Bardeen-Cooper-Schrieffer (BCS) superconductor resulted, in a certain region of parameters, in undamped time oscillations of the order parameter. However it was later \cite{dzero} realized that for system sizes larger than the superconducting coherence length the quench can excite finite momentum states. This results in spatial inhomogeneities of the order parameter that make the time oscillations unstable. Exact results
in a one dimension quantum spin-chain that is driven from paramagnetic to ferromagnetic \cite{spinchainexact} confirm this picture.

Despite these advances there is not yet a comprehensive theoretical framework to describe quantitatively most of these phenomena.
The recent introduction of the holographic principle, also called the (Anti de Sitter / Conformal field theory) AdS/CFT correspondence \cite{adscft}, in this context \cite{starinets,chesler,eke,bizon} has broaden considerably the theoretical tools to tackle non-equilibrium problems. 
In the context of holographic superconductivity \cite{hhh}, the problem we study here, there are already several studies that employ AdS/CFT techniques to describe the time evolution of the order parameter after a thermal \cite{murata,sonner,kam} or quantum \cite{us2012,basu} quench.
In these papers it was assumed that the order parameter was spatially uniform. This is a useful simplification since the gravity equations of motion depend only on two instead of three variables. However from the above discussion it is plausible that spatial inhomogeneities play a important role in the dynamic evolution of the order parameter. Indeed recent AdS/CFT calculations have shown \cite{liu} that a coupling to an axion field or a topological Chern-Simon term can induce a spontaneously breaking of translational invariance. Spatially inhomogeneous solutions of the gravity equations are also a crucial ingredient in the recent description of two dimensional turbulence \cite{liu1} by holographic techniques. The dynamics after a soft quench across a thermal or quantum critical point suggests \cite{basu2} that spatially inhomogeneous solutions might be stable. In the context of heavy ion collisions it was recently studied  the far-from equilibrium dynamics in the presence of spatial inhomogeneities \cite{keski1}.
It is therefore timely to ask whether the spontaneous breaking of translational symmetry can also be induced by a quench. Here we respond this question affirmatively.
We study the evolution of the order parameter of a holographic superconductor after a quench induced by turning on the source of the order parameter. Even though the initial state and the quench are spatially homogeneous we have observed that, for all quenches studied, the order parameter becomes spatially inhomogeneous for sufficiently long times. This spatially non-uniform solution has a lower free energy than the homogeneous one.
We start our analysis by introducing the gravity dual and working out the solutions of the equations of motion (EOM).
\section{The model and the boundary conditions}
 The starting action in the usual gravity dual of a holographic superconductor is \cite{hhh},
$
S=\int d^4x\sqrt{-g}[R-2\Lambda-\frac14F_{\mu\nu}F^{\mu\nu}-|\nabla\psi-iqA\psi|^2-m^2|\psi|^2]$
 where $\Lambda=-d(d-1)/2L^2$ is the cosmological constant while $d$ is the dimension of the boundary, $F_{\mu\nu}=\partial_\mu A_\nu-\partial_\nu A_\mu$ is the strength of the gauge field. The metric is an AdS Schwarzschild black hole,
 $
 ds^2=-f(r)dt^2+\frac{dr^2}{f(r)}+r^2(dx^2+dy^2)$
 with $f(r)=r^2/L^2(1-r_0^3/r^3)$, $r$ the bulk radial coordinate, $r_0$ the horizon position, and $x,y$ the boundary coordinates. Without loss of generality we set $q=1, L=1$.
 The temperature of the black hole is $
 T=\frac{3 r_0}{4 \pi}$.

We aim to find solutions that depend explicitly not only on time and the holographic coordinate $r$ but also on the spatial coordinate $x$ in the boundary, $\psi=\psi(t,r,x),~\psi^*=\psi^*(t,r,x)$,$~A=(A_t(t,r,x),A_r(t,r,x),A_x(t,r,x),0)$.
 However these functions are not gauge-invariant. In order to define gauge-invariant fields, we employ the following gauge transformations,
$
 \psi(t,r,x)=\rho(t,r,x)e^{i\varphi(t,r,x)},~\psi^*(t,r,x)=\rho(t,r,x)e^{-i\varphi(t,r,x)}$ and $
 A_i(t,r,x)=M_i(t,r,x)+\partial_i\varphi(t,r,x), ~~~i=t,r,x.$
The EOM resulting from the gravity Einstein equations for the gauge invariant fields $\rho$ and $M_i$ are,
\begin{widetext}
\be
&&\frac{\pa_x^2 M_t}{r^2 f}-\frac{2
   M_t \rho^2}{f}-\frac{\partial_{tx} M_x}{r^
   2 f}-\frac{2
   \partial_{t} M_r}{r}  -\partial_{tr} M_r+\frac{2
   \partial_{r} M_t}{r}+\partial_{r}^2 M_t=0,\label{eom1}\\
  && -\frac{f \partial_{x}^2 M_r}{r^2}+2 f
   M_r \rho^2+\frac{f
   \partial_{rx} M_x}{r^2}+\partial_{t}^2 M_r-\partial_{tr} M_t=0,\label{eom2}\\
  && f f' \partial_{x} M_r-f f'
   \partial_{r} M_x+f^2
   \partial_{rx} M_r-f^2
   \partial_{r}^2 M_x+2 f M_x
   \rho^2-\partial_{tx} M_t+\partial_{t}^2 M_x=0,\label{eom3}\\
   &&\rho \left(f
   (\frac{M_x^2}{r^2}+{m^2})
   +f^2
   M_r^2-M_t^2\right)-\frac{f \partial_{x}^2 \rho}{r^2}-f^2 \partial_{r}^2 \rho+\partial_{t}^2 \rho-\frac{f \left(r f'+2 f\right)
   \partial_{r} \rho}{r}=0,\label{eompsi1}\\
   &&\rho \left(-r^2 f f'
   M_r-r^2 f^2
   \partial_{r} M_r-2 r f^2
   M_r-f \partial_{x} M_x+r^2
   \partial_{t} M_t\right)-2 r^2 f^2
   M_r \partial_{r} \rho-2 f
   M_x \partial_{x} \rho  + 2 r^2
   M_t \partial_{t} \rho=0\label{eompsi2}.
\ee
\end{widetext}
We note that the phase $\varphi$ is automatically cancelled and that the last equation \eqref{eompsi2} is a linear combination of the first three equations, i.e., Eqs.\eqref{eom1},\eqref{eom2} and \eqref{eom3}. Therefore we have a well defined problem as there are four independent partial differential equations and four fields, $\rho, M_t, M_r$ and $M_x$.

In the limit of time independent fields, it is clear that only Eq.\eqref{eom1} and \eqref{eompsi1} survive as $M_r=0$ and $M_x=0$ are solutions to the above equations. However time dependent solutions require to turn  on $M_r$ and $M_x$ in order for the EOM to be gauge invariant and self-consistent.







We can now introduce the boundary conditions needed to solve the EOM.
Following the standard AdS/CFT dictionary, close to the boundary we impose,
\be
M_t&=&\mu(t,x)-\tilde{\rho}(t,x)/r+\dots, \label{muin}
\ee
and
$\rho = \rho_1(t,x)/r +\rho_2(t,x)/r^2+\dots,$
$ M_r = M_r^{(2)}(t,x)/r^2+\dots,$
$M_x = v(t,x)+ {\tilde J}(t,x)/r +\dots$
where we have set $m^2=-2$, $\tilde{\rho}$ is the charge density and $\mu$ is the chemical potential.
Before the quench we impose $\rho_1=0$ so that $\rho_2$ is the order parameter, \be \langle \mathcal O(x,t) \rangle \equiv \rho_2(x,t). \label{param} \ee
Since we do not consider the case of a finite super-current we can safely set ${\tilde J}=0$.
At the horizon we impose that $M_t=0$ and that the rest of functions have no singularities.
The next task is to define the thermal quench and to solve the EOM by a suitable numerical algorithm.\\
\begin{figure}
\begin{center}
\includegraphics[trim=0cm 5.5cm 2cm 4.0cm, clip=true,scale=0.4]{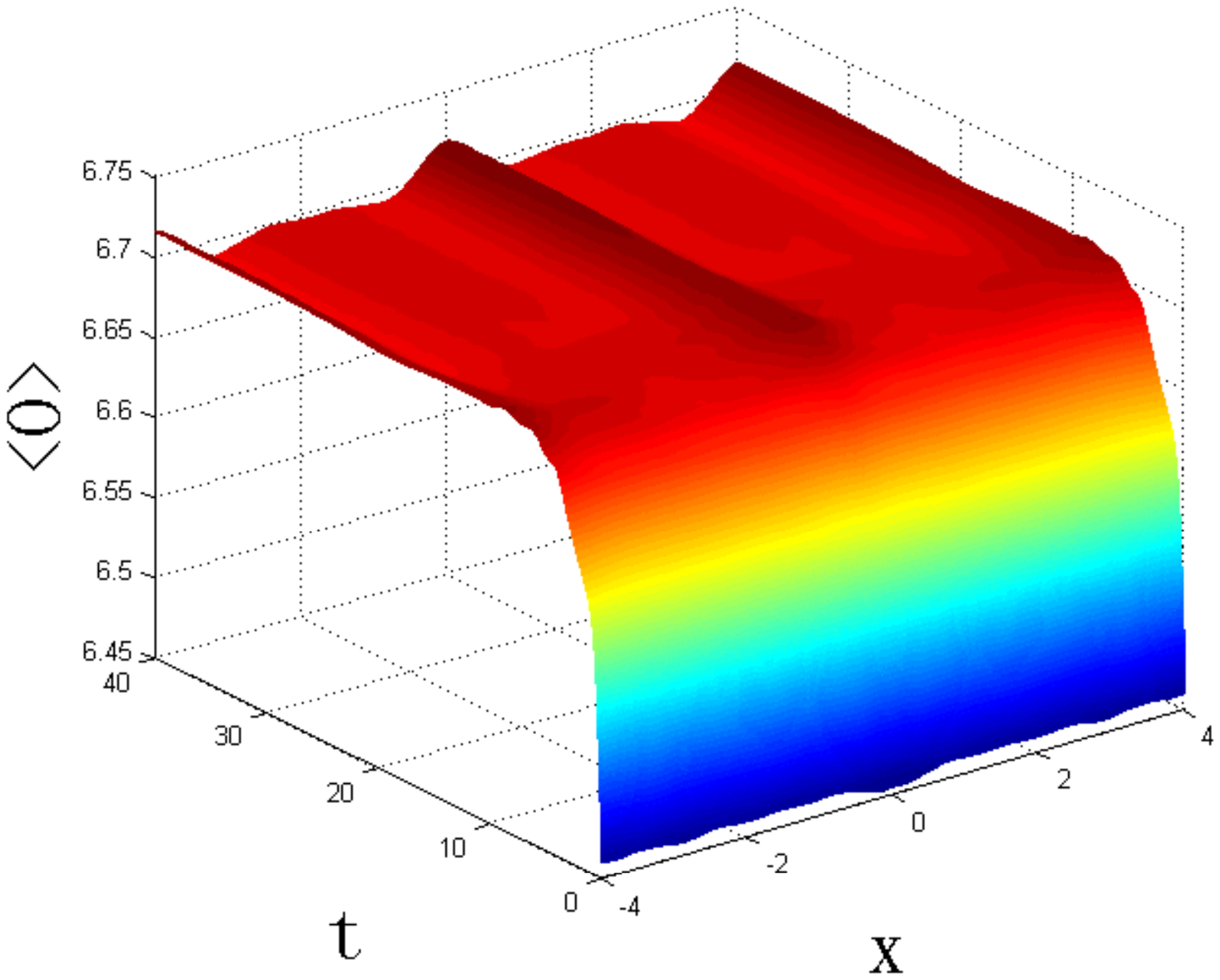}
\includegraphics[trim=4cm 5.5cm 2cm 5.5cm, clip=true,scale=0.45]{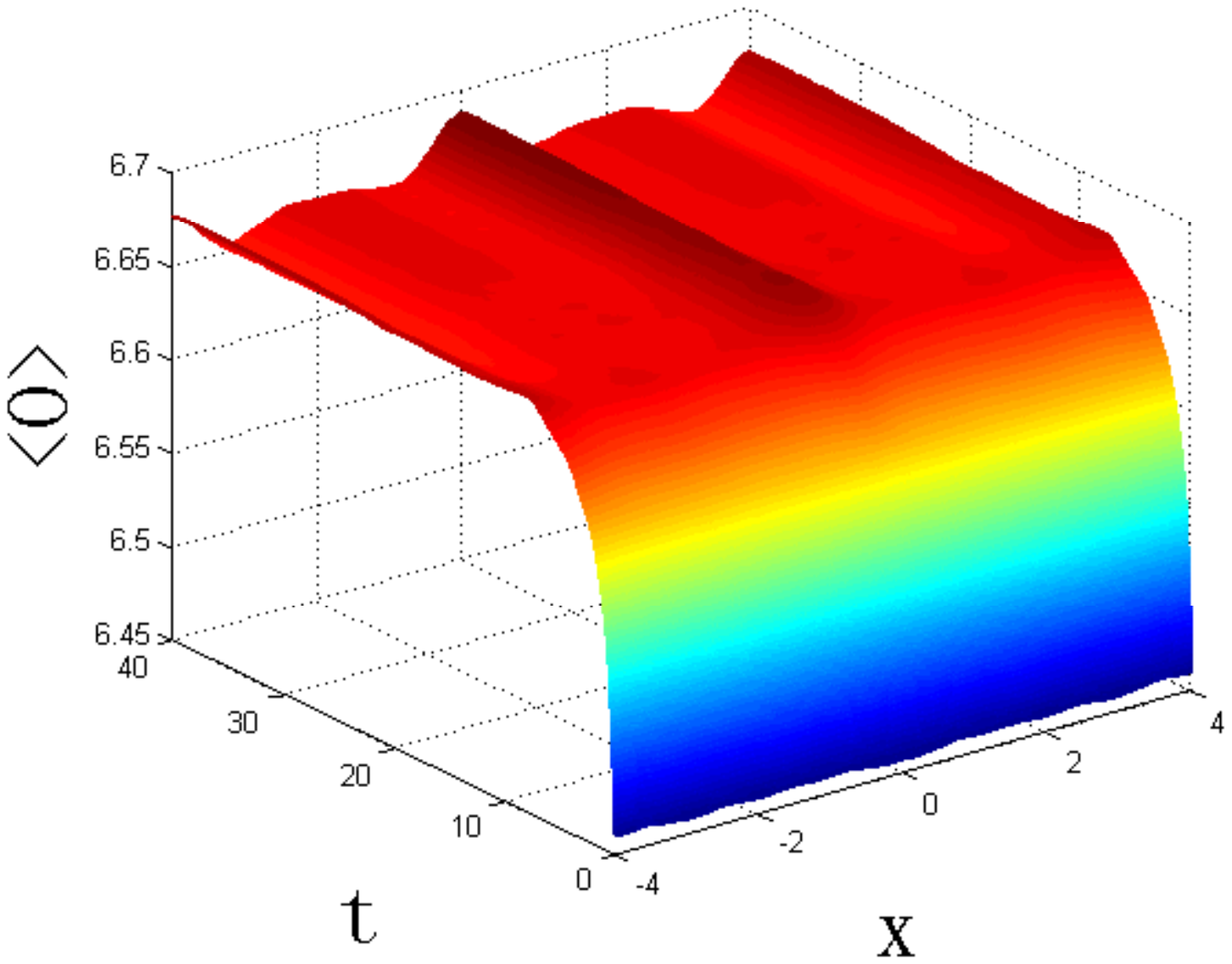}
\includegraphics[trim=0cm 6.5cm 2cm 5.5cm, clip=true,scale=0.45]{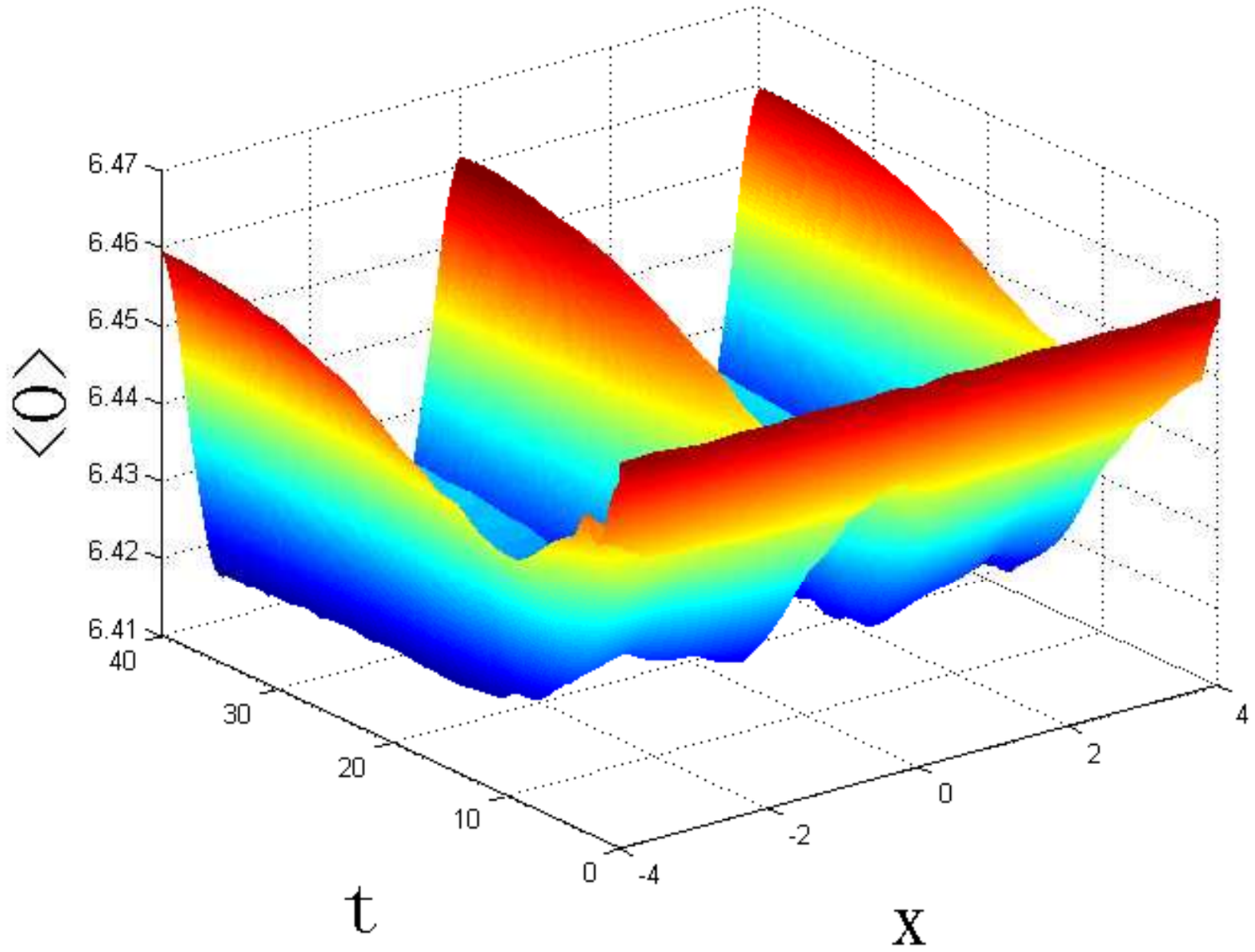}
\caption{Space and time dependence of the the order parameter $\langle \mathcal O(x,t) \rangle$ Eq. (\ref{param}) after the quench Eq.(\ref{que}) with $v=0.1$, $\mu=6$ and, from top to bottom, $J=1.5$, $J=1.2$ and $J=0.1$.}\label{fig1}
\end{center}
\end{figure}
\section{Definition of the quench and solution of the EOM}
We aim to study the time evolution of the order parameter $\langle \mathcal O(x,t) \rangle$ after a quench, namely, an abrupt change in the system. Following Ref.\cite{sonner,basu} we induce the quench by turning on the source in the expansion of scalar around the boundary, $\rho = \rho_1(t,x)/r +\rho_2(t,x)/r^2+\dots$ with \be \rho_1 = J \tanh(vt) \label{que}.\ee We note that with our choice of coordinates the order parameter is to a good approximation still given by $\rho_2(t,x)$. For $v \ll 1$ the source goes from zero to $J$ very quickly which we expect to excite the system to a far-from equilibrium state. We keep the chemical potential constant, $\mu=\mu_0$ and let the charge density vary. At $t=0$ we assume the system is described by a homogeneous and static solution of the EOM with $\mu(x,t=0)=\mu_0$.  The EOM evolves this initial homogeneous and static solution to a solution that verifies the above boundary condition for the scalar close to the boundary. We choose $v,J,\mu_0$ so that the system is always superfluid, namely, $\mu_0 > \mu_c \propto 1/T_c$ where in our quantization, $\mu_c=4.0636$ \cite{hhh}. We stay relatively close to $T_c$ so that the probe limit  that we employ is still a good approximation.
In order to solve the coupled partial differential equations
we used the spectral method \cite{spectral}. We discretize the EOM on a three
dimensional Chebyshev grid with $40$ points along the
$t$ direction and $20$ points along the $z =1/r$ direction, and up to $30$ points along the $x$ direction.
We study the time evolution for different values of $J,v$.
An important comment is in order. For technical reasons we do not have much flexibility to tune these parameters. If $v$ is very large then the perturbation is very slow so it is not really a quench. Moreover it will take a long time to observe any interesting effect. For $v$ too small the perturbation is very fast, a true quench, so we expect a relatively fine structure in the time and space evolution of the order parameter which cannot be resolved by the maximum number of points that we can simulate. More specifically we need that the coherence length which controls the spatial inhomogeneities be larger than the cutoff induced by the finite lattice spacing. That constraints the values of $J \leq 1$ and $v \sim 0.1$.
\section{The spatially inhomogeneous solution}
\begin{figure}
\begin{center}
\includegraphics[trim=2cm 7.5cm 2cm 12.5cm, clip=true,scale=0.9]{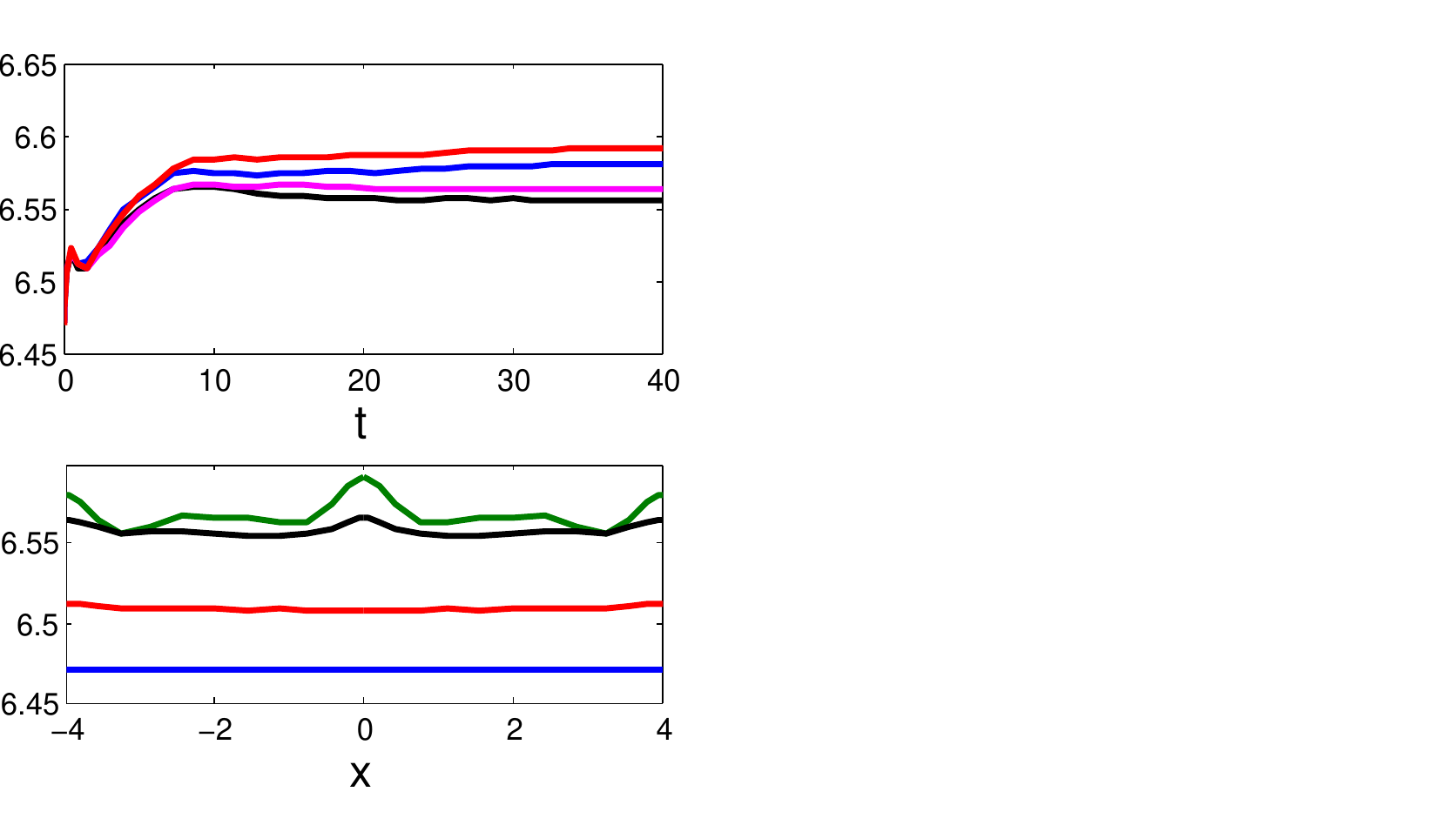}
\caption{Upper: The order parameter $\langle \mathcal O(t) \rangle$ for a quench $\mu=6,v=0.1,J=1.2$ as a function of time for different $x$'s: $x=0$ (red), $x=1.1322$ (pink),
$x= 3.247$ (black) and $x=4$ (blue). Lower: the order parameter $\langle \mathcal O(x) \rangle$, for the same quench but for different times,
$t=0$ (blue), $t=1$ (red)
$t=5$ (black), and $t=40$ (green).}\label{fig2}
\end{center}
\end{figure}

As was mentioned previously the problem of the dynamics of a holographic superconductor after a quench has already been investigated \cite{murata,sonner,kam} but in the limit of spatial homogeneity. Here we show that a thermal quench makes the order parameter spatially inhomogeneous at least for the abrupt changes of temperature that we explore in this paper. More importantly we provide compelling evidence that these solutions have a lower free
energy than the homogeneous ones. Results for the quench with $J=1$ and $v=0.1$ are shown in Fig.\ref{fig1} and Fig.\ref{fig2}. The time evolution is similar for different spatial points. However the spatial dependence strongly depends on time. For short times it is almost spatially homogeneous however, for longer times, spatial oscillations of growing amplitude are clearly observed.  The dependence on $x$ seems to be oscillatory which suggests that only few Fourier modes are excited by the quench. For smaller $v$ or larger $J$ we expect a more intricate pattern. However it would require a smaller lattice spacing which is beyond our numerical capabilities. The wavevectors $k_o$ of the oscillations of the order parameter is inversely proportional to the superconducting coherence length $k_o \sim n/\xi$ with $n$ an integer. For sufficiently strong quenches this coherence length does not have to correspond to the equilibrium one but rather to the one at which the evolution became non adiabatic \cite{zurek}. This is nothing but a consequence of the Kibble-Zurek mechanism.

 Physically this relatively simple oscillating pattern is an indication that the initially homogeneous order parameter decays into two states of finite and opposite momentum. Finally we stress that even though the temperature is well defined across the sample, namely the chemical potential $\mu$ is uniform, the order parameter $\langle \mathcal{O} \rangle$ still develops a spatial dependence that grows with time.

\begin{figure}
\begin{center}
\includegraphics[trim=1cm 5.5cm 0cm 5.5cm, clip=true,scale=0.4]{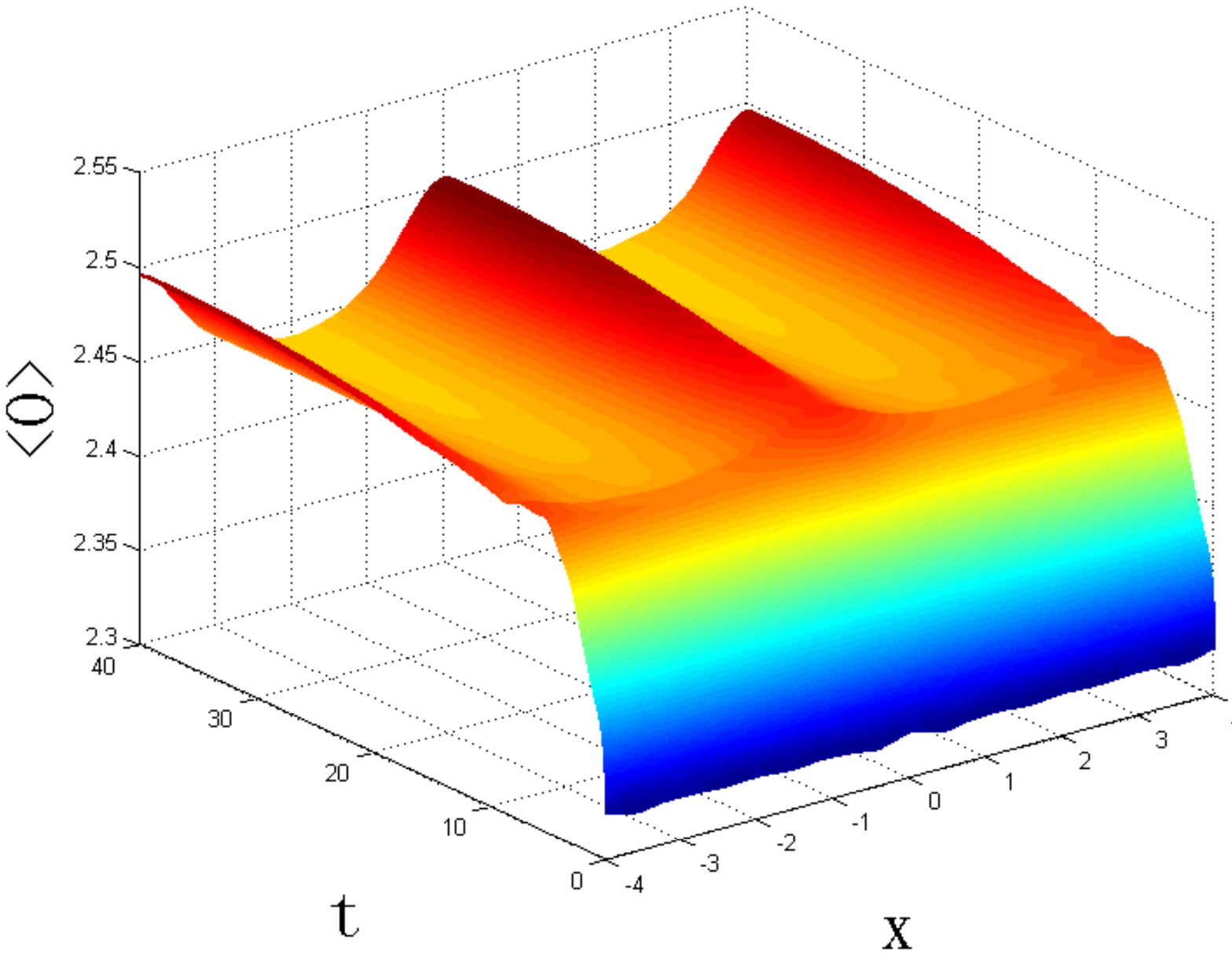}
\includegraphics[trim=0cm 5.5cm 1cm 5.5cm, scale=0.4,clip]{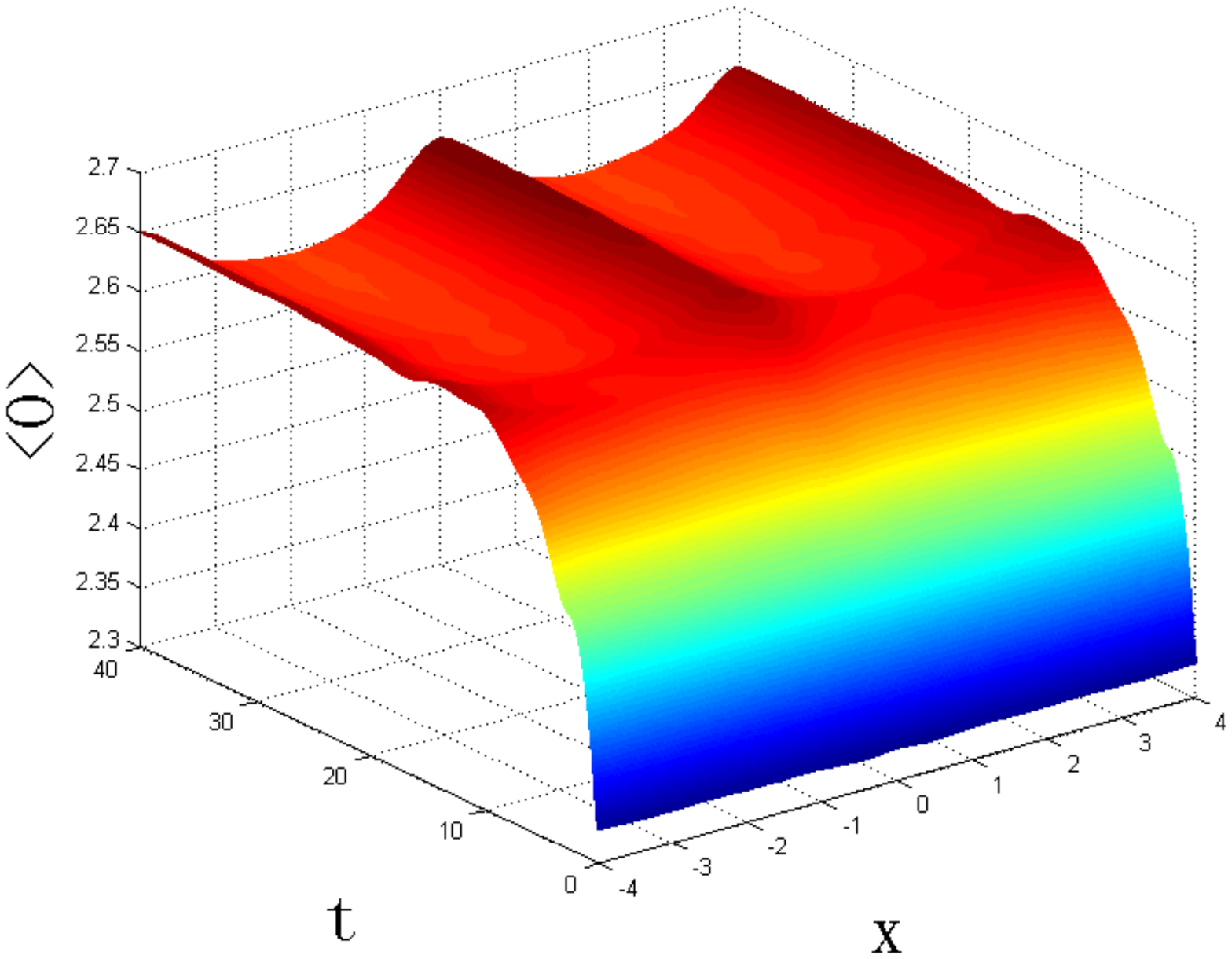}
\caption{Space and time dependence of the the order parameter $\langle \mathcal O(x,t) \rangle$ Eq. (\ref{param}) after the quench Eq.(\ref{que}) with $v=0.1$, $\mu=4.5$ and $J=0.8$ (upper), $J=1.2$ (lower). \label{fig3}
}
\end{center}
\end{figure}

These findings are consistent with those previously obtained \cite{emil} for weakly coupled superconductors in the condensed matter literature. Physically the spatial inhomogeneities \cite{emil} are a consequence of re-arrangements of the order parameter in space and time after a quench which are compatible with the conservation of energy and momentum \cite{emil}.
Similar results, depicted in Fig. \ref{fig3}, are observed for other quench parameters.
In summary, for all quenches, we see that the superconductor will eventually become spatially inhomogeneous. 
\section{Stability of the inhomogeneous solution}
Next we investigate whether these inhomogeneous solutions correspond to the physical state of minimum energy.
For that purpose we compare the free energy of the homogeneous and non-homogeneous solutions for long times around $t=t_f$ so that spatial inhomogeneities are more clearly observed.
\begin{figure}
\begin{center}
\includegraphics[trim=0cm 6.5cm 2cm 8.5cm, clip=true,scale=0.65]{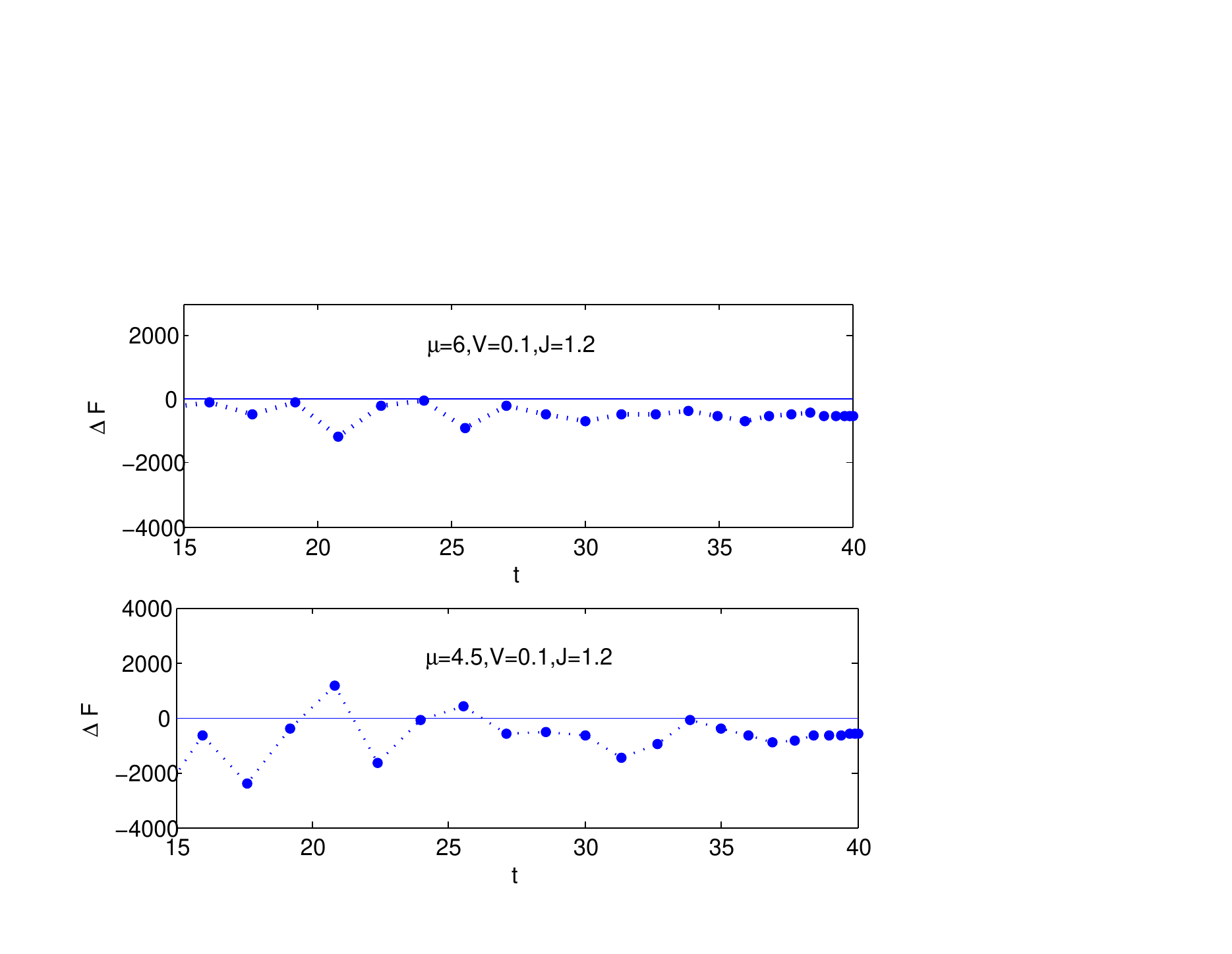}
\caption{The difference between the free energy Eq.(\ref{freee}) of the inhomogeneous and the homogeneous solution $\Delta F(t)$ as a function of time for two quenches. Interestingly the difference becomes clearly negative after the spatial inhomogeneities become substantial.}\label{fig5}
\end{center}
\end{figure}
The free energy $F=-TS_{os}+...$ is directly related to the on shell action $S_{os}$ \cite{Skenderis:2002wp}.
In order to evaluate $S_{os}$ it is convenient to integrate by parts and use the equations of motion which results in,
\be
S_{os}=\int_{z=0} d^3 x[\frac{h}{z^2}\rho \rho'-M_t M_t'+ h M_x M_x'+ h^2 M_rM_r'] - \nonumber \\ \int d^4 x [\frac{M_t^2}{h z^2} \rho-\frac{M_x^2 \rho^2}{z^2}+\frac{h}{z^2} M_r^2 \rho^2] \nonumber,
\ee
where $z = 1/r$, $h(z)=1-z^3$ and $'$ stands for the derivative with respect to z.
We work in the grand canonical ensemble characterized by a fixed $\mu$.
It is possible to show that boundary contributions have divergences coming from the scalar contribution. Fortunately this divergence can be removed by adding a counterterm. The resulting renormalized free energy is given by,
\begin{align}
F\propto \int_{z=0}d^3x[M_t M_t']+\int d^4x [\frac{M_t^2}{h z^2}\rho-\frac{M_x^2 \rho^2}{z^2}+\frac{h}{z^2}M_r^2\rho^2] \label{freee}
\end{align}
This expression is already suitable to compute the free energy for both homogeneous and inhomogeneous solutions at four different times which are close to $t_f$.
The results, depicted in Fig. \ref{fig5}, provide clear evidence that for all quenches the inhomogeneous solution has always a lower free energy.
This is a confirmation that, in general, thermal quenches not only make the order parameter time dependent but also space dependent. Therefore spatial inhomogeneity is an intrinsic ingredient in the dynamics of a strongly coupled superconductor after a homogeneous thermal quench.

In conclusion we have studied the time evolution of a holographic superconductor after abruptly turning on the source of the scalar field. For all the quenches studied the solution with the lowest free energy is spatially non-uniform. Time oscillations become unstable as spatial non homogeneities develop. To a good approximation the spatial dependence is a simple oscillatory function with an amplitude that increases with time until reaches, in the range of times studied, a constant value.

\acknowledgments
AMG acknowledges illuminating conversations with Hong Liu.
The authors thank Matthias Kaminski for fruitful discussions.
AMG acknowledge support from EPSRC, grant No. EP/I004637/1, FCT, grant PTDC/FIS/111348/2009 and
a Marie Curie International Reintegration Grant
PIRG07-GA-2010-268172. HQZ was supported by a Marie Curie International Reintegration Grant
PIRG07-GA-2010-268172. HBZ acknowledge support from FCT, grant PTDC/FIS/111348/2009, Marie Curie International Reintegration Grant
PIRG07-GA-2010-268172 and
the National Natural Science Foundation of China (under Grant No. 11205020).

\end{document}